\documentclass[12pt]{article}
\usepackage[letterpaper, margin=1in]{geometry}
\usepackage{times}
\usepackage{graphicx}
\usepackage{sectsty}
\allsectionsfont{\normalsize}
\usepackage{authblk}

\usepackage{amsmath, amsthm, amsfonts, amssymb}
\usepackage{arydshln}
\usepackage{enumerate}
\usepackage{array}
\usepackage{graphicx}
\usepackage{rotating}
\usepackage{tikz}
\usetikzlibrary{shapes.geometric, patterns}
\usetikzlibrary{shadings,intersections}
\usetikzlibrary{decorations.pathreplacing, angles, quotes}
\usetikzlibrary{arrows, decorations.markings}
\usetikzlibrary{plotmarks}

\theoremstyle{plain}

\begin{document}

\title{\normalsize\bf Emergent hierarchy through conductance-based node constraints}

\author[1, 2, 4]{C. Tyler Diggans}
\author[2,3]{Jeremie Fish}
\author[2,3]{Erik M. Bollt}

\affil[1]{Air Force Research Laboratory: Information Directorate, Rome, NY}
\affil[2]{Clarkson Center for Complex Systems Science (C$^3$S$^2$), Clarkson University}
\affil[3]{Department of Electrical and Computer Engineering, Clarkson University}
\affil[4]{Department of Physics, Clarkson University} 
\date{\vspace{-5ex}}
\maketitle
\thispagestyle{empty}
\pagestyle{empty}

\begin{abstract}\normalsize
The presence of hierarchy in many real-world networks is not yet fully explained. Complex interaction networks are often coarse-grain models of vast modular networks, where tightly connected subgraphs are agglomerated into nodes for simplicity of representation and feasibility of analysis. The emergence of hierarchy in growing complex networks may stem from one particular property of these ignored subgraphs: their graph conductance. Being a quantification of the main bottleneck of flow on the subgraph, all such subgraphs will then have a specific structural limitation associated with this scalar value. This supports the consideration of heterogeneous degree restrictions on a randomly growing network for which a hidden variable model is proposed based on the basic \textit{rich-get-richer} scheme. Such node degree restrictions are drawn from various probability distributions, and it is shown that restriction generally leads to increased measures of hierarchy, while altering the tail of the degree distribution. Thus, a general mechanism is provided whereby inherent limitations lead to hierarchical self-organization.
\end{abstract}

\section{Introduction}
Many real-world networks display both a measure of hierarchy~\cite{Clauset08} and a scale-free (SF) powerlaw form for some portion of the degree distribution. However, recent work~\cite{Broido19} has reminded us that the rich-get-richer SF model was only meant as a first approximation~\cite{Barabasi09}. In particular, the scale-free portion of the degree distribution does not necessarily extend into the region where the hubs are found\cite{Holme19}, indicating a potential relationship between the failings of the SF model and the observation of hierarchy. We shed light on this relationship by considering the consequences of the fact that many complex networks are coarse-grained models of more intricate, yet highly modular interaction networks, where the finer grained details are either computationally prohibitive or unknown. 

As an illustrative example to guide the discussion, we consider the extreme case of a social network, where each node represents a person. In turn, each person can be represented as a complex brain network, making the social network a course-grained model. In such models, the properties of the highly connected subgraphs that are being agglomerated into nodes are often ignored for simplicity. One property of the ignored sub-graph should not be overlooked and is highly relevant to transport and information flow: \textit{graph conductance}. The conductance of a network has been referred to as a quantification of the main bottleneck to the flow over the graph~\cite{Chung96}, meaning this scalar value may naturally place a limit on the potential degree of the agglomerated node. Furthermore, being a structural property, such degree limits would likely be heterogeneous and in fact follow a statistical distribution related to the domain application. 

As is often the case for details of these ignored subgraphs, the conductance values of the agglomerated nodes may not be known; even so, there are often proxy measurements that can quantify the effects produced by this fundamental restriction. For example, extrapolating from data on primate societies and the sizes of their neocortex, Robin Dunbar posited that human beings are likely capable of on average tracking a network of approximately $150$ human relationships~\cite{Dunbar92}. This result was supported by archaeological evidence of human tribe sizes along with the natural formation of divisions in large corporations. This is a clear example of the type of proxy value that might associate the conductance of the neural network of the neocortex of a specific node in a social network to that node's maximum allowable degree. 

A randomly growing network (RGN) model is presented that incorporates a node attribute as a hidden variable, which we call the node's \textit{internal conductance}. Prior work has explored Peer-to-Peer (P2P) networks with a homogeneous limit and has shown that imposing such hard limits on the degree of a rich-get-richer model leads to an exponential cutoff in the otherwise scale-free degree distribution~\cite{Guclu09}. On the other end, a specific mapping from the fine detailed structure of brain networks to social capability might one day be obtainable, but here we merely use this as a motivational assumption for the conceptual framwork and instead seek general results. In particular, we show that such restrictions lead to an increase of hierarchy in the resulting networks as measured by common measures such as the Global Reaching Centrality (GRC)~\cite{MVV2012} and the Random Walk Hierarchy ($H_{RW}$) measure~\cite{Czegel15}. 

For simplicity of presentation, we assume the true internal conductance values of the subgraphs can be mapped to an integer valued proxy measurement like Dunbar's number, which limits the degree that a node can sustain. Thus we focus on the choice for the distribution of these values. Although normal distributions are often used in assessing human capabilities such as with the Intelligence Quotient (IQ), this normality is often imposed on the data. Many such measures of human potential may actually follow a heavy tailed distribution~\cite{Psych2012}, though the mapping from a normally disributed network conductance to the proxy values may well be corrupted by social growth processes more generally. As such, we make no assumption on normality, and we compare the results for internal conductance values drawn from both localized and heavy tail distributions. The P2P network case is then incorporated into the present model by the choice of the dirac distribution. In fact, for the more localized distributions, such as Poisson and Binomial, we find similar results to the P2P hard limit case when the average internal conductance is relatively large. As these restrictions grow more severe, however, the measures of hierarchy separate in the different cases and the measures are reduced proportional to the variation due to the increased potential for hub formation. As may then be expected, the heavy tailed distributions also show the trend of increased hierarchy, though the effect is muted by the unbounded potential for hub formation regardless of the many restricted degree nodes.    

We begin by briefly defining graph conductance, however, we stress that the existence of the property of conductance is the only relevant fact, which informs the general concept supporting the proposed model. The actual values of that property are not considered in this model. Instead, a proxy measurement of maximum allowable degree is used in its place, leaving this mapping as a potential future direction of research. The hidden variable model is then described, while incorporating a review of the classic SF rich-get-richer model~\cite{Barabasi99}. The alterations to this traditional model include the assignment of a hidden node attribute and the implementation of a node dependent degree limit. We then define two measures of hierarchy that are used in the presentation of results for various distributions of the hidden variable. Following the presentation of results, we provide some explanation of how these restrictions lead to higher measures of hierarchy by alluding to the definitions of the measures themselves. We also show how the tail of the degree distributions are altered by this choice and provide some interesting future work in this direction.
\section{Preliminaries}
\subsection{Conductance}
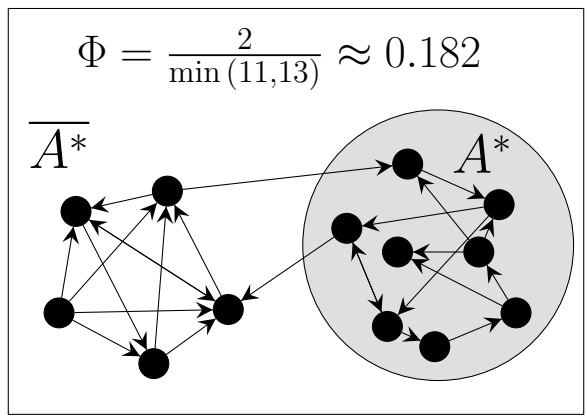
\begin{figure}[ht!]
\centering
\begin{tikzpicture}[scale=0.9]
\tikzset{myptr/.style={decoration={markings,mark=at position 1 with {\arrow[scale=2,>=stealth]{>}}},postaction={decorate}}}
\draw[fill=gray!25] (4.35,0) circle(2cm);
\node[draw, fill, circle] at (3,0.25) (A1) {};
\node[draw, fill, circle] at (5.5,-1) (B1) {};
\node[draw, fill, circle] at (4.95,-0.1) (C1) {};
\node[draw, fill, circle] at (3.6,-1.2) (D1) {};
\node[draw, fill, circle] at (3.75, -0.1) (E1) {};
\node[draw, fill, circle] at (5.25,0.6) (F1) {};
\node[draw, fill, circle] at (3.9, 1.2) (G1) {};
\node[draw, fill, circle] at (4.3, -1.5) (H1) {};

\node[draw, fill, circle] at (-1.25,-1) (A2) {};
\node[draw, fill, circle] at (-1,0.5) (B2) {};
\node[draw, fill, circle] at (0.35,0.8) (C2) {};
\node[draw, fill, circle] at (1.25,-0.95) (D2) {};
\node[draw, fill, circle] at (0.15, -1.75) (E2) {};

\draw[myptr] (B1) to (C1);
\draw[myptr] (C1) to (E1);
\draw[myptr] (D1) to (A1);
\draw[myptr] (C1) to (G1);
\draw[myptr] (B1) to (E1);
\draw[myptr] (C1) to (F1);
\draw[myptr] (D1) to (H1);
\draw[myptr] (A1) to (D1);
\draw[myptr] (H1) to (B1);
\draw[myptr] (G1) to (F1);
\draw[myptr] (F1) to (A1);
\draw[myptr] (F1) to (D1);

\draw[myptr] (A1) to (D2);
\draw[myptr] (C2) to (G1);

\draw[myptr] (A2) to (B2);
\draw[myptr] (A2) to (C2);
\draw[myptr] (A2) to (D2);
\draw[myptr] (A2) to (E2);
\draw[myptr] (B2) to (D2);
\draw[myptr] (B2) to (E2);
\draw[myptr] (C2) to (B2);
\draw[myptr] (D2) to (B2);
\draw[myptr] (E2) to (C2);
\draw[myptr] (E2) to (D2);
\draw[myptr] (D2) to (C2);

\node at (5,1.35) {\LARGE$A^*$};
\draw (-2,-2.5) rectangle (6.5,3.5);
\node at (-1.25,1.5) {\LARGE$\overline{A^*}$};
\node at (2,2.75) {\Large$\Phi=\frac{2}{\min{(11,13)}}\approx 0.182$};
\end{tikzpicture}
\caption{An illustration of how to compute the conductance of a graph. The set $A^*$ is the subset of vertices which minimizes $\Phi(G)$. The relatively small conductance value indicates there is a significant bottleneck in the graph.}
\label{Dumbbell}
\end{figure}

The conductance of a graph is an isoperimetric number associated with that graph. The original isoperimetric problem was to identify which curve of fixed length would enclose the largest volume, and subsequent versions were analogues of this question for manifolds and graphs seeking to associate the surface area of the object to that of the volume. Given a graph $G=(V,E)$, where $V$ is the set of nodes and $E$ the set of edges, the conductance of $G$ has been defined as 
\begin{equation}
\Phi = \Phi(G) = \min_{A\subset V}\frac{|E\left(A,\overline{A}\right)|}{\min\left\{vol\left(A\right),vol\left(\overline{A}\right)\right\}},
\label{GRC}
\end{equation}
where $\overline{A}$ is the complement of $A$; $E(A,B)$ denotes the set of edges with one node in set $A$ and the other node in set $B$; and $vol(A)$ is the sum of the out degrees of all nodes in set $A$. The optimal solution set $A^*$ has a minimal number of links from itself to the rest of the graph, while also preferring $A^*$ and its complement to have a comparable number of internal links. In this way, a quantitative measure for what has been described as the main bottleneck of the graph is provided~\cite{Chung96}. See Figure~\ref{Dumbbell} for an illustration of a solution set.

\subsection{Hierarchy}
The historical definition and understanding of the term hierarchy is complicated, and may best be described in terms of category theory; but, the measurement of hierarchy in complex networks has recently reached a reasonable, albeit fractured, concensus~\cite{Krackhardt94},~\cite{Luo11},~\cite{MVV2012}~\cite{Murtra2013},~\cite{Czegel15}. Of the widely known measures of hierarchy, two are considered in this paper: the Global Reaching Centrality (GRC)~\cite{MVV2012} and the Random Walk Hierarchy measure ($H_{RW}$)~\cite{Czegel15}. Perhaps the most versatile measure of hierarchy, presented in 2012, is the GRC. It is intuitively based on the definition of hierarchy as a \textit{heterogeneous distribution of centrality}, where the ideal network would have few nodes with large centrality values and relatively many nodes with smaller values. The main advantage of this measure is the inclusion of undirected and weighted networks with simple alterations to the formulae. The GRC is described in terms of the set of reaching centralities of node $i$, denoted $Cr(i)$, for all $i\in 1,2,\dots,N$. More specifically, the GRC is defined as the average difference of these values from the maximum reaching centrality $Cr_{max}$ as:
\begin{equation}
GRC = \frac{\sum_i^N{Cr_{max} - Cr(i)}}{N-1}.
\label{GRC}
\end{equation} 
This same formula can be adapted by defining how $Cr(i)$ is computed depending on whether the graph is directed or weighted. In the presently considered case of directed unweighted graphs, $Cr(i)$ is simply the proportion of nodes that are reachable along directed edges from node $i$. It is clear from the definition ~(\ref{GRC}) that having few nodes with large centrality will result in many large contributions to the overall sum. In this way, the GRC measures the spread the distribution of centrality, in particular promoting distributions with exponentially decreasing histograms of $Cr$ values. This captures the essence of tree-like networks, without explicitly requiring tree-like structure. For example, a flower graph~\cite{Rozenfeld06}, which has no discernible tree-like structure and many cycles, is still very hierarchical by the GRC measure.   

Having identified a potential issue with the widely accepted GRC measure, the Random Walk hierarchy measure ($H_{RW}$) was introduced in 2015 in order to penalized structures that were technically tree-like, though are not noticeably hierarchical, such as chains or star graphs that actually attain large GRC values. This alternate measure is formulated through simulated diffusion of decaying random walkers. It is a measure of the spread of the stationary distribution of these random walkers ($\textbf{p}^{stat}$) and can be computed in closed form by
\begin{equation}
H_{RW}=\sqrt{N\sum_{i=1}^N\left(p_i^{stat}\right)^2-1};
\end{equation} 
with the $i^{\textrm{th}}$ element of the stationary distribution being given by
\begin{equation}
\begin{aligned}
p_i^{stat}&=\frac{e^{1/\lambda}-1}{N}\left(e^{1/\lambda}\mathbb{I}_N - \hat{\bf{T}}\right)^{-1} \hat{\bf{T}}\cdot \hat{\bf{1}}\\
&=\frac{e^{1/\lambda}-1}{N}\sum_{n=1}^\infty{\left(e^{-1/\lambda}\hat{\bf{T}}\right)^n\hat{\bf{1}}};
\end{aligned}
\end{equation} 
where $\hat{\bf{T}}$ is a stochastic transition matrix computed by setting the probability of transition from node $j$ to node $i$ such that $P(j\rightarrow i)\propto\frac{1}{k_j^{in}k_i^{out}}$, $\hat{\bf{1}}$ is the vector of  ones, and $\lambda$ is a parameter that represents the characteristic distance for which the weight of the random walker is decreased to $e^{-1}$, which was shown to be optimally set to $\lambda=4$. The series representation given is used for approximations when matrix inversion becomes too computaionally expensive. This measure targets a heterogeneity in the distances between pairs of points, and we chose it to illustrate the limits of our model's seeming success for extreme restrictions under the GRC.

\section{The Internal Conductance Model}
\begin{figure}[ht!]
\centering
\begin{tikzpicture}[scale=0.9]
\tikzset{myptr/.style={decoration={markings,mark=at position 1 with {\arrow[scale=3,>=stealth]{>}}},postaction={decorate}}}
\node[draw, circle] at (0,0) (A) {$a$};
\node[draw, circle] at (2,0) (B) {$b$};
\node[draw, circle] at (-1,-2) (C) {$c$};
\node[draw, circle] at (1,-2.5) (D) {$d$};
\node[draw, circle] at (2.5,-3.5) (E) {$e$};
\node[draw, circle] at (-0.75,-3.75) (F) {$f$};
\node[draw, circle] at (3,-1.75) (G) {$g$};
\node[draw, circle] at (-2.5,-2.75) (H) {$h$};
\node[draw, circle] at (0,-4.5) (I) {$i$};
\node[draw, circle] at (5,-3) (J) {$j$};
\node[draw, circle] at (2,-5.5) (K) {$k$};
\draw[myptr] (B) to (A);
\draw[myptr] (C) to (A);
\draw[myptr] (D) to (A);
\draw[myptr] (C) to (B);
\draw[myptr] (D) to (B);
\draw[myptr] (E) to (D);
\draw[myptr] (E) to (B);
\draw[myptr] (F) to (D);
\draw[myptr] (F) to (B);
\draw[myptr] (G) to (A);
\draw[myptr] (G) to (B);
\draw[myptr] (H) to (A);
\draw[myptr] (H) to (C);
\draw[myptr] (I) to (A);
\draw[myptr] (I) to (D);
\draw[myptr] (J) to (B);
\draw[myptr] (J) to (D);
\draw[myptr] (K) to (J);
\draw[myptr] (K) to (D);
\end{tikzpicture}
\caption{An example of a Barab\'{a}si-Albert style SF net using the parameter $m=2$ to avoid tree-like growth. The proposed internal conductance model follows this construction, except that each node has a specific degree restriction (theoretically obtained from its hidden subgraph's conductance). Nodes are taken out of the pool of potential links from new nodes once their degree reaches its \textit{internal conductance} value, which is an abstraction of the conductance into degree limit.}
\label{BA}
\end{figure}
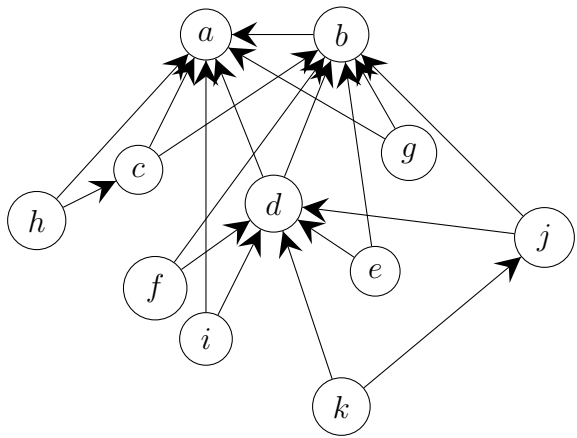

In order to focus our attention on the main concepts presented, we chose to alter the basic Albert-Barab\'{a}si \textit{rich-get-richer} model using $m=2$, which was shown to create Scale-Free (SF) powerlaw degree distributions~\cite{Barabasi99}. In this basic model, one node is added in each generation and that node is connected to $m$ previously generated nodes; these nodes are chosen at random but are weighted by their current degrees. One specific network that might be generated by this model is shown in Figure~\ref{BA} for reference.

We alter the SF model by assigning an \textit{internal conductance} value to each node. This variable is drawn from a positive valued distribution that may be chosen based on assumptions concerning the application at hand. In fact, since we are using unweighted networks for simplicity of presentation, we chose integer-valued distributions that can be thought of as a proxy or an abstraction of the conductance values to the effective restriction on links that the conductance value would impart. Thus, during the random growth process, once a node's degree meets its assigned internal conductance, that node is no longer considered in the selection of parents for new nodes. For example, referring back to Figure~\ref{BA}, if node $a$ was assigned an internal conductance of $5$ upon creation, after the current generation it would no longer be eligible to parent newly added nodes. However, it is important to note that node $a$ would still likely be an ancestor of many more new nodes due to its high relative number of children at this point in the process. This feature of increased likelihood of becoming an ancestor but not a direct parent is important to showing why measures of hierarchy increase as the restrictions are imposed.

\section{Results}
To date, the majority of proposed generative hierarchical network models have been prescriptive, while many real-world networks are inherently dependent on growth dynamics and some form of preferential attachment. The intuition created by the work of Robin Dunbar on primate community sizes led to the proposition that the ignored subgraphs' conductance values were playing a key role in the structure of real-world networks. By the description of conductance as the bottleneck of information flow, this property of the subgraphs imply the fundamental existence of a restriction on the allowable flow through any given node. Furthermore, there is no reason to assume homogeneity in the conductance values for these agglomerated subgraphs.

While the term hierarchy remains difficult to define generally, the most widely used metric is the GRC; furthermore, the concept of a measure of the heterogeneity of the centrality of nodes, which is central to the definition of the GRC, is considered. In the original SF model, due to the directed links originating from new nodes, the newest nodes are the most likely to attain the maximum reach centrality, $Cr_{max}$. This reversal in which the newest nodes are thought of as parents in terms of directed edges, leads to networks with quite low GRC values in general. Regardless, we find that as distributions with lower average internal conductance values are used, the nodes that would have become hub-like (having large in-degree) are taken out of the pool of potential links from newer nodes. As mentioned previously, these nodes that reach their maximum in-degree may not be direct descendents of any additional new nodes, but they do have a much higher likelihood of becoming distant descendents of many future nodes. This is the essence of how hierachy emerges in this model.

\begin{figure}[ht!]
\centering
\begin{tabular}{cccc}
\includegraphics[width=2.75in]{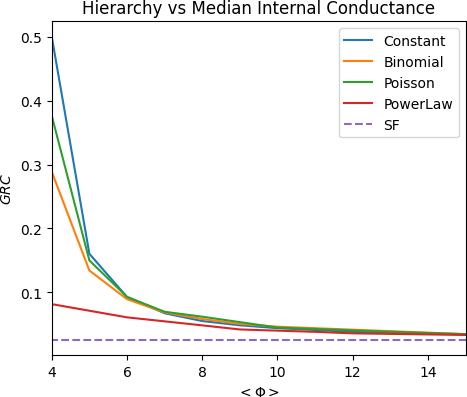}
& & &
\includegraphics[width=2.75in]{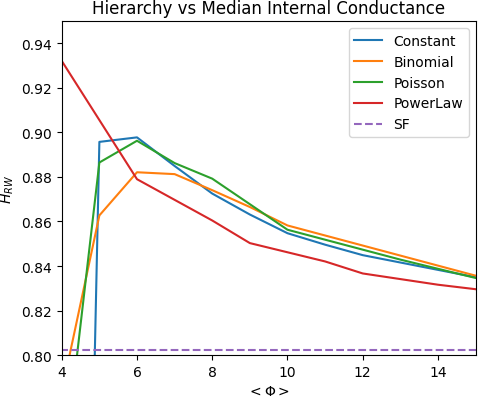}\\
(a) & & & (b)\\
\end{tabular}
\caption{Measures of (a) GRC and (b) $H_{RW}$ against the median conductance value of the nodes averaged over $1000$ trials for graph sizes of $N=2000$ nodes, i.e. 2 million nodes, using python networkx. Curves represent different choices for the distribution of internal conductance values with variation in parameters. Both hierarchy measures increase as the restrictions are imposed through either lowering the mean values or shifting the $xmin$ variable within the powerlaw python package. The trend breaks down in the Random Walk Hierarchy measure, $H_{RW}$, which was designed specifically to penalize trees with small spreads such as chains or $m$-ary trees for small $m$. The Scale Free values are included as a dashed line to which all measures converge for sufficiently large conductance values, i.e. $\langle\Phi\rangle>>N$.}
\label{Distros}
\end{figure}

Figure~\ref{Distros} shows the change in values of (a) the GRC and (b) the Random Walk hierarchy measure, $H_{RW}$, as a function of the median internal conductance values for networks of size $N=2000$ averaged over $1000$ trials. While the influence of this alteration is subtle at large average conductance values, as the restrictions get more severe, the measures of hierarchy increase rapidly with an acceleration dependent on the variation in the distribution, until the trend breaks down. Especially for $H_{RW}$, when the average degree limit reaches around $6$, the ideal spreading between a chain and star graph is attained for the formulation of that measure. Any additional restrictions leads to lower measures of hierarchy as the network becomes more chain-like.

\begin{figure}[ht!]
\centering
\includegraphics[width=4in]{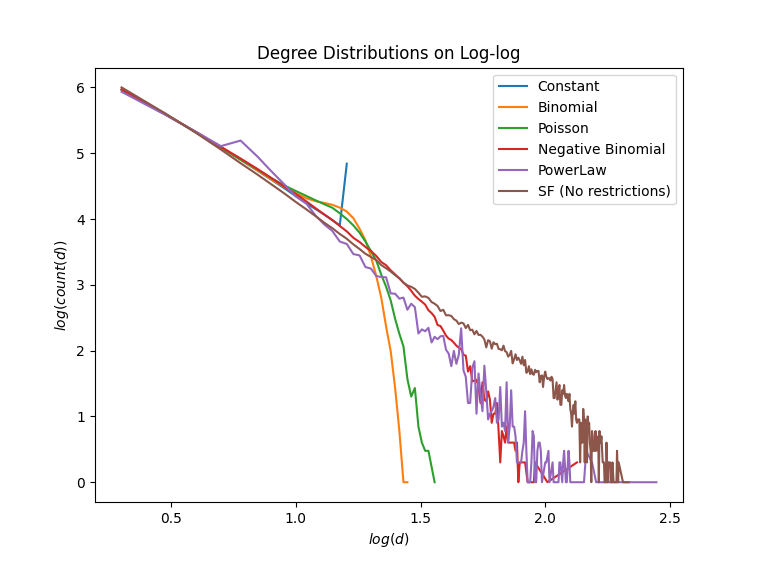}
\caption{A log-log plot of the degree distributions for $1000$ trials of graph sizes $N=2000$ nodes, i.e. 2 million nodes, resulting from various choices of internal conductance distributions with comparable mean values near $\langle\Phi\rangle\approx 16$.}
\label{Degrees}
\end{figure}

In general, the rich-get-richer growth process is essentially unaltered for nodes with small degrees, until some of the internal limits are attained. This quenching of degree growth leads to two properties of their degree distributions. There is a bunching up of degrees near the average conductance value for localized distributions followed by an exponential cut-off of the otherwise SF degree distribution, similar to what was observed in P2P networks~\cite{Guclu09}. The negative binomial distribution was included in Figure~\ref{Degrees} in order to provide an additional example of the effects caused by a heavy tailed distribution for the internal conductance values. The general trend being that those distributions which result in more SF-like degree distributions are less influenced in terms of the average increase in hierarchy due to the higher probablity of hub formation.

\section{Conclusions}
We have presented a paradigm for understanding why real-world networks tend to display increased hierarchical structure over what is expected from purely scale-free growth. This framework only assumes that many networks are coarse-grained models of more intricate networks, and that the ignored properties of the underlying structure impact the growth process in subtle ways. Nothing in the physical world has true limitless potential. Networks of information flow are restricted by the processing speed of each node, and rivers are restricted by the height of their banks. As such, when considering randomly growing networks with the intention of real world models, there is no evidence to support limitless growth of a node's degree. By imposing homogeneous restrictions on the degrees of the nodes, we find the expected exponential cut-off in the degree distribution that was shown in P2P networks. In addition, we find this aleration of the tail of the degree distribution to be associated with an increase measure of hierarchy. Future work will explore how the growth process itself may impact the development of conductance values under a use-it-or-lose-it growth assumption in an effort to identify the correct distribution to use for this model to effectively reproduce real-world network structures. We will also explore how the subtle alteration of the tail of the degree distribution may help explain the break down of the SF model for accurate reflections of reality.  
\section*{Acknowledgments}
CTD gratefully acknowledges funding from the Air Force Office of Scientific Research in support of this work.

\bibliography{ms}
\bibliographystyle{unsrt}
\end{document}